\documentclass[conference]{IEEEtran}
\usepackage{cite}
\usepackage{amsmath,amssymb,amsfonts}
\usepackage{multirow}
\usepackage{tabularx}
\usepackage{caption}
\usepackage[subrefformat=parens]{subcaption}
\usepackage{algorithmic}
\usepackage{graphicx}
\usepackage{textcomp}
\usepackage{xcolor}
\usepackage{bm}

\def\Hline{\noalign{\hrule height 0.4mm}}

\newcommand{\ie}{\textit{i.e.}}
\newcommand{\eg}{\textit{e.g.}}
\newcommand{\etal}{\textit{et~al.}}

\def\tr{\mathrm{T}}
\def\hr{\mathrm{H}}

\def\eqdef{\overset{\mbox{\fontsize{6pt}{0pt}\selectfont def}}{=}}
\def\eqc{\overset{\mbox{\fontsize{6pt}{0pt}\selectfont c}}{=}}

\def\ComplexGaussian{\mathcal{N}_{\mathbb{C}}}

\def\A{\mathbf{A}}
\def\B{\mathbf{B}}

\def\D{\mathbf{D}}

\def\G{\mathbf{G}}
\def\H{\mathbf{H}}
\def\I{\mathbf{I}}

\def\Q{\mathbf{Q}}

\def\U{\mathbf{U}}
\def\V{\mathbf{V}}
\def\W{\mathbf{W}}
\def\X{\mathbf{X}}
\def\Y{\mathbf{Y}}
\def\Z{\mathbf{Z}}

\def\e{\mathbf{e}}

\def\g{\mathbf{g}}

\def\q{\mathbf{q}}
\def\r{\mathbf{r}}

\def\x{\mathbf{x}}
\def\y{\mathbf{y}}
\def\z{\mathbf{z}}

\def\emalgo{{${}_{\mbox{\tiny EM}}$}}

\allowdisplaybreaks[4]

\begin{document}

\title{Fast Multichannel Source Separation
Based on Jointly Diagonalizable Spatial Covariance Matrices}

\author{
\IEEEauthorblockN{
Kouhei Sekiguchi\IEEEauthorrefmark{1}\IEEEauthorrefmark{2}
\ \ \
Aditya Arie Nugraha\IEEEauthorrefmark{1}
\ \ \
Yoshiaki Bando\IEEEauthorrefmark{3}
\ \ \
Kazuyoshi Yoshii\IEEEauthorrefmark{1}\IEEEauthorrefmark{2}
}\\
\IEEEauthorblockA{
\IEEEauthorrefmark{1}Center for Advanced Intelligence Project (AIP), RIKEN, Tokyo 103-0027, Japan
\\
Email: \{kouhei.sekiguchi, adityaarie.nugraha, kazuyoshi.yoshii\}@riken.jp
}
\IEEEauthorblockA{
\IEEEauthorrefmark{2}Graduate School of Informatics, Kyoto University, Kyoto 606-8501, Japan
}
\IEEEauthorblockA{
\IEEEauthorrefmark{3}National Institute of Advanced Industrial Science and Technology (AIST), Tokyo, 135-0064, Japan
\\
Email: y.bando@aist.go.jp
}
}

\maketitle
\fussy

\begin{abstract}
This paper describes a versatile method 
 that accelerates multichannel source separation methods 
 based on full-rank spatial modeling.
A popular approach to multichannel source separation
 is to integrate a spatial model with a source model
 for estimating the spatial covariance matrices (SCMs) 
 and power spectral densities (PSDs) of each sound source
 in the time-frequency domain.
One of the most successful examples of this approach
 is multichannel nonnegative matrix factorization (MNMF) 
 based on a full-rank spatial model and a low-rank source model.
MNMF, however, is computationally expensive and often works poorly 
 due to the difficulty of estimating the unconstrained full-rank SCMs.
Instead of restricting the SCMs to rank-1 matrices
 with the severe loss of the spatial modeling ability
 as in independent low-rank matrix analysis (ILRMA),
 we restrict the SCMs of each frequency bin 
 to jointly-diagonalizable but still full-rank matrices.
For such a fast version of MNMF,
 we propose a computationally-efficient 
 and convergence-guaranteed algorithm
 that is similar in form to that of ILRMA.
Similarly,
 we propose a fast version of a state-of-the-art speech enhancement method
 based on a deep speech model and a low-rank noise model.
Experimental results showed that 
 the fast versions of MNMF and the deep speech enhancement method
 were several times faster and performed even better 
 than the original versions of those methods, respectively.
\end{abstract}


\section{Introduction}

Multichannel source separation plays a central role 
 for computational auditory scene analysis.
To make effective use of an automatic speech recognition system
 in a noisy environment, for example,
 it is indispensable to separate speech signals from noise-contaminated signals.
A standard approach to multichannel source separation
 is to use a non-blind method (\eg, beamforming and Wiener filtering) 
 based on the spatial covariance matrix (SCM) 
 of a target source (\eg, speech) and those of the other sources (\eg, noise).
To use beamforming for speech enhancement,
 deep neural networks (DNNs) are often used 
 for classifying each time-frequency bin 
 into speech or noise~\cite{Narayanan2013, Erdogan2016, Heymann2016}.
The performance of such a supervised approach, however, 
 is often considerably degraded in an unseen environment.
In this paper we thus focus on general-purpose blind source separation (BSS)
 and its extension for environment-adaptive semi-supervised speech enhancement.

\begin{figure}[t]
\centering
\includegraphics[width=.95\linewidth]{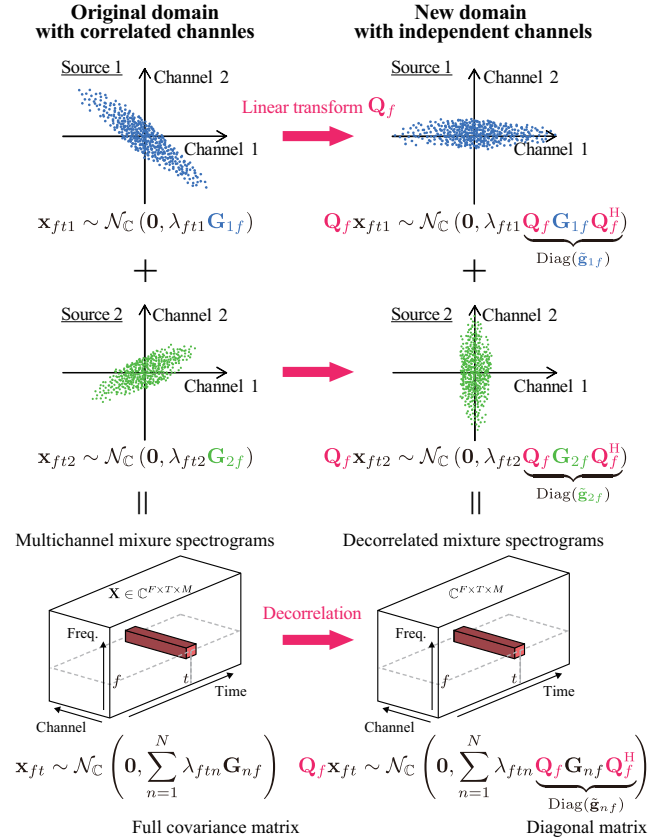}
\caption{The full-rank spatial model 
for the \textit{correlated} channels of the original data is equivalent
to the diagonal spatial model 
for the \textit{independent} channels of the decorrelated data.} 
\label{fig:joint_diagonalization}
\vspace{-2mm}
\end{figure}

The goal of BSS is to estimate 
 both a mixing process and sound sources from observed mixtures.
To solve such an ill-posed problem,
 one can take a statistical approach based on a \textit{spatial model} 
 representing a sound propagation process
 and a \textit{source model}
 representing the power spectral densities (PSDs) of each source.
Duong \etal~\cite{Duong2010} pioneered this approach
  by integrating a full-rank spatial model 
 using the frequency-wise full-rank SCMs of each source
 with a source model assuming the source spectra to follow complex Gaussian distributions.
 We call it as full-rank spatial covariance analysis (FCA) in this paper as in \cite{Ito2018}.
To alleviate the frequency permutation problem of FCA,
 multichannel nonnegative matrix factorization (MNMF)
 that uses an NMF-based source model 
 for representing the co-occurrence and low-rankness of frequency components 
 has been developed~\cite{Ozerov2010, Arberet2010, Sawada2013}.
Such a low-rank source model, however, does not fit speech spectra. 
In speech enhancement,
 a semi-supervised approach that uses as source models 
 a DNN-based speech model (deep prior, DP) trained from clean speech data
 and an NMF-based noise model learned on the fly
 has thus recently been investigated
 (called MNMF-DP)~\cite{Bando2018, Sekiguchi2018, Leglaive2019}.

The major drawbacks common to these methods based on the full-rank SCMs
 are the high computational cost 
 due to the repeated heavy operations (\eg, inversion) of the SCMs
 and the difficulty of parameter optimization
 due to the large degree of freedom (DOF) of the spatial model.
Kitamura \etal~\cite{Kitamura2016} thus proposed a constrained version of MNMF
 called independent low-rank matrix analysis (ILRMA)
 that restricts the SCMs to \textit{rank-1} matrices.
Although ILRMA is an order of magnitude faster 
 and practically performed better than MNMF,
 it suffers from the severe loss of the spatial modeling ability.
Ito \etal~\cite{Ito2018}
 proposed a fast version of FCA
 that restricts the SCMs of each frequency bin 
 to \textit{jointly-diagonalizable} matrices.
For parameter estimation,
 an expectation-maximization (EM) algorithm 
 with a fixed-point iteration (FPI) method
 was proposed, but its convergence was not guaranteed.
 
In this paper 
 we propose a versatile convergence-guaranteed method 
 for estimating the jointly-diagonalizable SCMs of the full-rank spatial model
 and its application to 
 FCA, MNMF, and MNMF-DP
 called FastFCA, FastMNMF, and FastMNMF-DP, respectively,
 where FastMNMF has an intermediate ability of spatial modeling between MNMF and ILRMA.
As shown in Fig.~\ref{fig:joint_diagonalization},
 while all channels are correlated in the original spectrograms,
 they are independent in the linearly-transformed spectrograms
 obtained by applying a diagonalizer to each frequency bin.
MNMF for the original \textit{complex} spectrograms
 is thus equivalent to computationally-efficient nonnegative tensor factorization (NTF)
 for the independent \textit{nonnegative} PSDs of the transformed spectrograms.
To estimate such a diagonalizer (linear transform),
 we use an iterative projection (IP) method
 in a way similar to independent vector analysis (IVA)~\cite{Ono2011}
 that estimates a demixing matrix.
The resulting algorithm based on iterations of NTF and IP
 is similar in form to that of ILRMA based on iterations of NMF and IP.

One of the important contributions of this paper 
 is to improve existing decomposition methods
 by joint diagonalization of covariance matrices.
This idea was first discussed 
 for an ultimate but computationally-prohibitive extension of NTF
 called correlated tensor factorization (CTF)~\cite{Yoshii2018a}
 based on multi-way full-rank covariance matrices,
 resulting in a fast version of CTF
 called independent low-rank tensor analysis (ILRTA)~\cite{Yoshii2018}.
While ILRTA was used for single-channel BSS 
 based on jointly-diagonalizable \textit{frequency} covariance matrices,
 in this paper we focus on multi-channel BSS 
 based on jointly-diagonalizable \textit{spatial} covariance matrices.
Since NTF and IP are used in common for parameter optimization,
 the proposed FastMNMF can be regarded as a special case of ILRTA.

\section{Multichannel Source Separation}
\label{sec:conventional_method}

This section reviews existing multichannel source separation methods based on a full-rank spatial model,
 \ie, full-rank spatial covariance analysis (FCA)~\cite{Duong2010}
 based on an unconstrained source model,
 MNMF~\cite{Sawada2013}
 based on an NMF-based source model,
 and its adaptation to speech enhancement called MNMF-DP~\cite{Sekiguchi2018}
 based on a DNN-based speech model and an NMF-based noise model.

\subsection{Full-Rank Spatial Model}
\label{sec:spatial_modeling}

\subsubsection{Model Formulation}

Suppose that $N$ sources are observed by $M$ microphones. 
Let $\X = \{\x_{ft}\}_{f,t=1}^{F,T} \in \mathbb{C}^{F \times T \times M}$
 be the observed multichannel complex spectra,
 where $F$ and $T$ are the number of frequency bins
 and that of frames, respectively.
Let $\x_{ftn} = [x_{ftn1}, \cdots, x_{ftnM}]^\tr \in \mathbb{C}^{M}$
 be the image of source $n$
 assumed to be circularly-symmetric complex Gaussian distributed as follows:
\begin{align}
 \x_{ftn} \sim \ComplexGaussian \left(\mathbf{0}, \lambda_{ftn} \G_{nf}\right),
 \label{eq:p_x_ftn}
\end{align}
where $\lambda_{ftn}$ is the PSD of source $n$ at frequency $f$ and time $t$, 
 $\G_{nf}$ is the $M \times M$ positive definite full-rank SCM of source $n$ at frequency $f$.
Using the reproductive property of the Gaussian distribution,
 the observed spectrum $\x_{ft} = \sum_{n=1}^N \x_{ftn}$ is given by
\begin{align}
 \x_{ft} \sim \ComplexGaussian\!\left( \mathbf{0}, \sum_{n=1}^{N} \lambda_{ftn} \G_{nf} \right).
 \label{eq:p_x_ft}
\end{align}
\fussy

Given the mixture spectrum $\x_{ft}$ 
 and the model parameters $\G_{nf}$
 and $\lambda_{ftn}$,
 the posterior expectation of the source image $\x_{ftn}$ is obtained
 by multichannel Wiener filtering (MWF):
\begin{align}
 \x_{ftn}
 = \mathbb{E}[\x_{ftn} | \x_{ft}]
 = \Y_{ftn} \Y_{ft}^{-1} \x_{ft} ,
 \label{eq:sep_mwf}
\end{align}
where $\Y_{ftn} \!\eqdef\! \lambda_{ftn} \G_{nf}$
 and $\Y_{ft} \!\eqdef\! \sum_{n=1}^N \!\! \Y_{ftn}$.

\subsubsection{Parameter Estimation}

Our goal is to estimate 
 the parameters $\G = \{\G_{nf}\}_{f,n=1}^{F,N}$
 and $\bm\Lambda = \{\lambda_{ftn}\}_{f,t,n=1}^{F,T,N}$
 that maximize the log-likelihood given by Eq.~(\ref{eq:p_x_ft}):
\begin{align}
\log p({\X} | \G, \bm\Lambda) 
 \eqc
 - \sum_{f,t=1}^{F,T} 
 \left(
 \mathrm{tr}\!\left( \X_{ft} \Y_{ft}^{-1} \right) 
 + \log\!\left|\Y_{ft}\right|
 \right),
\label{eq:log_likelihood}
\end{align}
 where $\X_{ft} \!\eqdef\! \x_{ft}\x_{ft}^\hr$.
In this paper
 we use a majorization-minimization (MM) algorithm~\cite{Sawada2013}
 that iteratively maximizes a lower bound of Eq.~(\ref{eq:log_likelihood}).
As in \cite{Yoshii2018a,Yoshii2018},
 the closed-form update rule of $\G$ was recently found to be given by
\begin{align}
\A_{nf} &\eqdef \textstyle\sum_{t=1}^T \lambda_{ftn} \Y_{ft}^{-1} \X_{ft} \Y_{ft}^{-1},
\\
\B_{nf} &\eqdef \textstyle\sum_{t=1}^T \lambda_{ftn} \Y_{ft}^{-1},
\\
\G_{nf}
&\leftarrow
\B_{nf}^{-1} \left( \B_{nf} \G_{nf} \A_{nf} \G_{nf}  \right)^{\frac{1}{2}}.
\label{eq:update_G}
\end{align}

\subsection{Source Models}
\subsubsection{Unconstrained Source Model}

The unconstrained model directly uses $\bm\Lambda$ as free parameters.
Using the MM algorithm,
 the multiplicative update (MU) rule of $\bm\Lambda$
 is given by
\begin{align}
 \lambda_{ftn} \leftarrow \lambda_{ftn} 
 \sqrt{\frac{\mathrm{tr}\!\left( \G_{nf}\Y_{ft}^{-1} \X_{ft} \Y_{ft}^{-1} \right) }
 {\mathrm{tr}\!\left(\G_{nf} \Y_{ft}^{-1}\right)}}.
\end{align}

\subsubsection{NMF-Based Source Model}

If the PSDs $\{\lambda_{ftn}\}_{f,t=1}^{F,T}$ of a source $n$ (\eg, noise and music)
 have low-rank structure, 
 the PSDs can be factorized as follows \cite{Sawada2013}:
\begin{align}
 \lambda_{ftn} = \sum_{k=1}^K w_{nkf} h_{nkt},
 \label{eq:lambda_NMF}
\end{align}
where $K$ is the number of bases, $w_{nkf} \ge 0$ is the magnitude of basis $k$ of source $n$ at frequency $f$, 
 and $h_{nkt} \ge 0$ is the activation of basis $k$ of source $n$ at time $t$.
Using the MM algorithm\cite{Nakano2010},
 the MU rules of $\W$ and $\H$ are given by
\begin{align}
 w_{nkf}
 &\leftarrow
 w_{nkf} \sqrt{\frac{\sum_{t=1}^T h_{nkt}
 \,\mathrm{tr}\! \left(\G_{nf} \Y_{ft}^{-1} \X_{ft} \Y_{ft}^{-1}\right)}{\sum_{t=1}^T h_{nkt} 
 \,\mathrm{tr}\! \left(\G_{nf} \Y_{ft}^{-1}\right)}},
 \label{eq:update_w}
 \\
 h_{nkt}
 &\leftarrow
 h_{nkt} \sqrt{\frac{\sum_{f=1}^F w_{nkf}
 \,\mathrm{tr}\! \left(\G_{nf} \Y_{ft}^{-1} \X_{ft} \Y_{ft}^{-1}\right)}{\sum_{f=1}^F w_{nkf}
 \,\mathrm{tr}\! \left(\G_{nf} \Y_{ft}^{-1}\right)}}.
 \label{eq:update_h}
\end{align}

\subsubsection{DNN-Based Source Model}

To represent the complicated characteristics of the PSDs $\{\lambda_{ftn}\}_{f,t=1}^{F,T}$ of a source $n$ (\eg, speech),
 a deep generative model can be used as follows\cite{Bando2018}:
\begin{align}
  \lambda_{ftn} = 
  u_{nf} v_{nt} [\bm\sigma^2_{\bm\theta}(\z_{nt})]_f
  \label{eq:lambda_DSP}
\end{align}
where $\bm\sigma^2_{\bm\theta}(\cdot)$ 
 is a nonlinear function (DNN) with parameters $\bm\theta$
 that maps a latent variable $\z_{nt} \in \mathbb{R}^D$ 
 to a nonnegative spectrum $\r_{nt} \!\eqdef\! \bm\sigma^2_{\bm\theta}(\z_{nt}) \in \mathbb{R}_+^{F}$ at each time $t$,
 $[\cdot]_f$ indicates the $f$-th element of a vector,
 $u_{nf} \ge 0$ is a scaling factor at frequency $f$,
 and $v_{nt} \ge 0$ is an activation at time $t$.

To update the latent variables $\Z_n = \{\z_{nt}\}_{t=1}^T$,
 we use Metropolis sampling.
A proposal $\z_{nt}^\mathrm{new} \sim \mathcal{N}(\z_{nt}^\mathrm{old}, \epsilon \I)$ is accepted 
 with probability ${\rm min}\left(1, \gamma_{nt} \right)$,
 where $\gamma_{nt}$ is given by
\begin{align}
\log  \gamma_{nt} 
&= - \sum_{f=1}^F \left(\frac{1}{\lambda_{ftn}^\mathrm{new}} - \frac{1}{\lambda_{ftn}^\mathrm{old}} \right)
 \mathrm{tr}\! \left(\G_{nf} \Y_{ft}^{-1} \X_{ft} \Y_{ft}^{-1}\right)
 \nonumber \\
 &\quad
 - \sum_{f=1}^F \left(\lambda_{ftn}^\mathrm{new} - \lambda_{ftn}^\mathrm{old} \right)
 \mathrm{tr}\! \left(\G_{nf} \Y_{ft}^{-1} \right),
 \label{eq:acceptance_rate}
\end{align}
where $\lambda_{ftn}^\mathrm{new} = u_{nf} v_{nt} [\bm\sigma_{\bm\theta}^2(\z_{nt}^\mathrm{new})]_f$,
$\lambda_{ftn}^\mathrm{old} = u_{nf} v_{nt} [\bm\sigma_{\bm\theta}^2(\z_{nt}^\mathrm{old})]_f$.
In practice, we update $\Z_n$ several times without updating $\Y_{ft}$
 to reduce the computational cost of calculating $\Y_{ft}^{-1}$.

In the same way as the NMF-based source model,
 the MU rules of $\U$ and $\V$ are given by
\begin{align}
 u_{nf} 
 &\leftarrow
 u_{nf} \sqrt{\frac{\sum_{t=1}^T v_{nt} r_{ntf} 
 \mathrm{tr}\! \left(\G_{nf} \Y_{ft}^{-1} \X_{ft} \Y_{ft}^{-1}\right)}{\sum_{t=1}^T v_{nt} r_{ntf} 
 \mathrm{tr}\!\left(\G_{nf} \Y_{ft}^{-1}\right)}} ,\\
 v_{nt} 
 &\leftarrow
 v_{nt} \sqrt{\frac{\sum_{f=1}^F u_{nf} r_{ntf} 
 \,\mathrm{tr}\! \left(\G_{nf} \Y_{ft}^{-1} \X_{ft} \Y_{ft}^{-1}\right)}{\sum_{f=1}^F u_{nf} r_{ntf} 
 \,\mathrm{tr}\!\left(\G_{nf} \Y_{ft}^{-1}\right)}}.
\end{align}

\subsection{Integration of Spatial and Source Models}

\subsubsection{Full-Rank Spatial Covariance Analysis}
\label{sec:fca}

FCA\cite{Duong2010} is obtained 
 by integrating the full-rank spatial model and the unconstrained source model.
While the EM algorithm was originally used in \cite{Duong2010},
 in this paper we use the MM algorithm
 expected to converge faster as in \cite{Sawada2013,Yoshii2018}.

\subsubsection{Multichannel NMF}
\label{sec:mnmf}

MNMF\cite{Sawada2013} is obtained 
 by integrating the NMF-based source model into FCA.

\subsubsection{MNMF with a Deep Prior}
\label{sec:mnmf_dp}

MNMF-DP\cite{Sekiguchi2018} specialized for speech enhancement
 is obtained by integrating the full-rank spatial model
 and the DNN- and NMF-based source models
 representing speech and noise sources, respectively.
Assuming a source indexed by $n=1$ corresponds to the speech,
$\lambda_{ft1}$ and $\lambda_{ft(n \ge 2)}$
are given by Eq.~(\ref{eq:lambda_DSP}) and Eq.~(\ref{eq:lambda_NMF}), respectively.

\section{Fast Multichannel Source Separation}
\label{sec:proposed_method}

This section proposes the fast versions 
 of FCA, MNMF, and MNMF-DP based 
 on the joint diagonalizable SCMs.

\subsection{Jointly Diagonalizable Full-Rank Spatial Model}
\subsubsection{Model Formulation}

To reduce the computational cost of the full-rank spatial model, 
 we put a constraint that the SCMs $\{\G_{nf}\}_{n=1}^N$
 can be jointly diagonalized as follows:
\begin{align}
 \Q_f \G_{nf} \Q_f^{\hr} = \mathrm{Diag}(\tilde{\g}_{nf}),
 \label{eq:diagonalize}
\end{align}
where $\Q_f = [\q_{f1}, \cdots, \q_{fM}]^\hr \in \mathbb{C}^{M \times M}$ is a non-singular matrix 
 called a \textit{diagonalizer}
 and $\tilde{\g}_{nf} = [\tilde{g}_{nf1}, \cdots, \tilde{g}_{nfM}] \in \mathbb{R}_+^{M}$ is a nonnegative vector.
The observed spectrum $\x_{ft}$ is projected into a new space 
 where the elements of the projected spectrum $\Q_f \x_{ft}$ are all independent
 (Fig.~\ref{fig:joint_diagonalization}).
 
\subsubsection{Parameter Estimation}
 
Our goal is to jointly estimate $\Q$, $\tilde\G$, and $\mathbf{\Lambda}$
 that maximize the log-likelihood given 
 by substituting Eq.~(\ref{eq:diagonalize}) into Eq.~(\ref{eq:p_x_ft})
 as follows:
\begin{align}
  &\!\!
  \log p(\X |\Q, \tilde\G, \bm\Lambda) 
  \nonumber\\
  &\!\!
  = \sum_{f,t=1}^{F,T}
  \log \ComplexGaussian\!\left({\x_{ft}} \Bigg| \mathbf{0}, \sum_{n=1}^N \lambda_{ftn} \Q_f^{-1} \mathrm{Diag}(\tilde{\g}_{nf}) \Q_f^{-\hr} \right)
  \nonumber\\
  &\!\!
  \eqc
  \sum_{f,t,m=1}^{F,T,M}
  \!\!
  \left( -\frac{ \tilde{x}_{ftm} }{\tilde{y}_{ftm}} - \log \tilde{y}_{ftm} \right) + T \sum_{f=1}^F \log \left| \Q_{f} \Q_{f}^{\hr} \right|,
  \label{eq:log_likelihood_new}
\end{align}
where 
 $\tilde{\x}_{ft} {=} \mathrm{Diag}(\Q_f \X_{ft} \Q_f^\hr) {=}  |\Q_f \x_{ft}|^{\circ 2}$,
 $|\cdot|^{\circ 2}$ indicates the element-wise absolute square,
 and $\tilde{\y}_{ft} = \sum_{n=1}^N \lambda_{ftn} \tilde{\g}_{nf}$.

Since Eq.~(\ref{eq:log_likelihood_new}) has the same form 
 as the log-likelihood function of IVA \cite{Ono2011},
 $\Q_f$ can be updated 
 by using the convergence-guaranteed iterative projection (IP) method
 as follows:
\begin{align}
    \V_{fm} &\eqdef \frac{1}{T} \sum_{t=1}^T \X_{ft} \tilde{y}_{ftm}^{-1}, \\
    \q_{fm} &\leftarrow (\Q_f \V_{fm})^{-1} \e_m , \\
    \q_{fm} &\leftarrow (\q_{fm}^{\hr} \V_{fm} \q_{fm})^{-\frac{1}{2}} \q_{fm},
    \label{eq:update_Q}
\end{align}
where $\e_m$ is a one-hot vector whose $m$-th element is 1.
A diagonalizer $\Q_f$ is estimated
 so that the $M$ components (\textit{channels}) of $\{\Q_f \x_{ft}\}_{f,t=1}^{F,T}$ 
 become independent.
In IVA~\cite{Ono2011} and ILRMA~\cite{Kitamura2016} under a determined condition ($M=N$),  
 a demixing matrix $\D_f$ is estimated
 so that the $M$ components (\textit{sources}) 
 of $\{\D_f \x_{ft}\}_{f,t=1}^{F,T}$ 
 become independent.
In any case,
 the characteristics of the components (\eg, low-rankness in the NMF-based source model)
 represented by $\{\tilde\y_{ft}\}_{f,t=1}^{F,T}$ are considered.
This implies that our method could work as fast as ILRMA 
 even in an underdetermined condition ($M < N$) 
 while keeping the full-rank spatial modeling ability.

Since the first term of Eq.~(\ref{eq:log_likelihood_new})
 is the negative Itakura-Saito (IS) divergence 
 between $\tilde{x}_{ftm}$ and $\tilde{y}_{ftm}$,
 the MU rule of $\tilde\G$ is given
 by using the MM algorithm for IS-NMF\cite{Nakano2010} as follows:
\begin{align}
    \tilde{g}_{nfm} \leftarrow 
    \tilde{g}_{nfm}
    \sqrt{
    \frac
    {\sum_{t=1}^{T} \lambda_{ftn} \tilde{y}_{ftm}^{-1} \tilde{x}_{ftm}\tilde{y}_{ftm}^{-1}}
    {\sum_{t=1}^{T} \lambda_{ftn} \tilde{y}_{ftm}^{-1}}
    }.
    \label{eq:update_g}
\end{align}

\subsection{Source Models}

\subsubsection{Unconstrained Source Model}

Using the MM algorithm for IS-NMF\cite{Nakano2010}, 
 the MU rule of $\bm\Lambda$ is given by
\begin{align}
    \lambda_{ftn} &\leftarrow \lambda_{ftn}
    \sqrt{\frac{\sum_{m=1}^M  \tilde{g}_{nfm} \tilde{y}_{ftm}^{-1} \tilde{x}_{ftm}\tilde{y}_{ftm}^{-1} } {\sum_{m=1}^M \tilde{g}_{nfm} \tilde{y}_{ftm}^{-1} }}.
    \label{eq:update_lambda}
\end{align}

\subsubsection{NMF-Based Source Model}

Similarly, 
 the MU rules of $\W$ and $\H$ included in Eq.~(\ref{eq:lambda_NMF})
 are given by
\begin{align}
    w_{nkf} &\leftarrow w_{nkf} \sqrt
    { \frac{\sum_{t,m=1}^{T, M} h_{nkt} \tilde{g}_{nfm} \tilde{y}_{ftm}^{-1} \tilde{x}_{ftm}\tilde{y}_{ftm}^{-1} } 
    {\sum_{t, m=1}^{T, M} h_{nkt} \tilde{g}_{nfm} \tilde{y}_{ftm}^{-1}} }, \\
    h_{nkt} &\leftarrow h_{nkt} \sqrt
    { \frac{\sum_{f,m=1}^{F, M} w_{nkf} \tilde{g}_{nfm} \tilde{y}_{ftm}^{-1} \tilde{x}_{ftm}\tilde{y}_{ftm}^{-1} } 
    {\sum_{f, m=1}^{F, M} w_{nkf} \tilde{g}_{nfm} \tilde{y}_{ftm}^{-1}} }.
\end{align}

\subsubsection{DNN-Based Source Model}

To update the latent variables $\Z_n$ included in Eq.~(\ref{eq:lambda_DSP}),
 we use Metropolis sampling.
A proposal $\z_{nt}^\mathrm{new} \sim \mathcal{N}(\z_{nt}^\mathrm{old}, \epsilon \I)$ is accepted 
 with probability ${\rm min}\left(1, \gamma_{nt} \right)$,
 where $\gamma_{nt}$ is given by
\begin{align}
    \log \gamma_{nt} = 
    & - \sum_{f, m=1}^{F,M} \left( \frac{\tilde{x}_{ftm}} {\lambda_{ftn}^{\rm{new}} \tilde{g}_{nfm} + \tilde{y}^{\neg n}_{ftm} } - \frac{\tilde{x}_{ftm}} {\lambda_{ftn}^{\rm{old}} \tilde{g}_{nfm} + \tilde{y}^{\neg n}_{ftm} }  \right) \nonumber \\
    & - \sum_{f, m=1}^{F,M} \log \frac{\lambda_{ftn}^{\rm{new}} \tilde{g}_{nfm} + \tilde{y}^{\neg n}_{ftm}}{\lambda_{ftn}^{\rm{old}} \tilde{g}_{nfm} + \tilde{y}^{\neg n}_{ftm}},
\end{align}
where $ \tilde{y}^{\neg n}_{ftm} \!\eqdef\! \sum_{n' \not= n}^N \lambda_{ftn'}\tilde{g}_{n'fm}$
 is a reconstruction without the component of source $n$.
As in the NMF-based source model,
 the MU rules of $\U$ and $\V$ are given by
\begin{align}
    u_{nf} &\leftarrow u_{nf} \sqrt{
        \frac{\sum_{t,m=1}^{T, M} v_{nt} r_{ntf} \tilde{g}_{nfm} \tilde{y}_{ftm}^{-1} \tilde{x}_{ftm}\tilde{y}_{ftm}^{-1}  }
        {\sum_{t,m=1}^{T, M} v_{nt} r_{ntf} \tilde{g}_{nfm} \tilde{y}_{ftm}^{-1}}
    }, \\
    v_{nt} &\leftarrow v_{nt} \sqrt{
        \frac{\sum_{f,m=1}^{F, M} u_{nf} r_{ntf} \tilde{g}_{nfm} \tilde{y}_{ftm}^{-1} \tilde{x}_{ftm}\tilde{y}_{ftm}^{-1}  }
        {\sum_{f,m=1}^{F, M} u_{nf} r_{ntf} \tilde{g}_{nfm} \tilde{y}_{ftm}^{-1} }
    }.
\end{align}

\subsection{Integration of Spatial and Source Models}

\subsubsection{FastFCA}
\label{sec:fast_fca}

The fast version of FCA is obtained 
 by integrating the jointly diagonalizable full-rank spatial model
 and the unconstrained source model.
While the EM algorithm with the FPI step was originally used in \cite{Ito2018},
 in this paper we use the MM algorithm 
 with the convergence-guaranteed IP step.

\subsubsection{FastMNMF}
\label{sec:fast_mnmf}

The fast version of MNMF is obtained 
 by integrating the NMF-based source model into FastFCA.

\subsubsection{FastMNMF-DP}
\label{sec:fast_mnmf_dp}

The fast version of MNMF-DP
 is obtained by integrating the jointly diagonalizable full-rank spatial model, 
 the DNN- and NMF-based source models representing speech and noise sources, respectively.

\section{Evaluation}

This section evaluates the performances and efficiencies
 of the proposed methods in a speech enhancement task.

\subsection{Experimental Conditions}

100 simulated noisy speech signals sampled at 16 kHz 
 were randomly selected from the evaluation dataset of CHiME3~\cite{Barker2017}.
These data were supposed 
 to be recorded by six microphones attached to a tablet device.
Five channels ($M = 5$) excluding the second channel behind the tablet were used.
The short-time Fourier transform
 with a window length of 1024 points ($F=513$) 
 and a shifting interval of 256 points was used.
To evaluate the performance of speech enhancement,
 the signal-to-distortion ratio (SDR) was measured \cite{Vincent2006, Raffel2014}.
To evaluate the computational efficiency, 
 the elapsed time per iteration for processing 8 sec data was measured
 on Intel Xeon W-2145 (3.70 GHz) or NVIDIA GeForce GTX 1080 Ti.

FastFCA (Section~\ref{sec:fast_fca}), 
 FastMNMF (Section~\ref{sec:fast_mnmf}), 
 and FastMNMF-DP (Section~\ref{sec:fast_mnmf_dp})
 based on the \textit{jointly diagonalizable} SCMs
 were compared with FCA (Section~\ref{sec:fca}), 
 MNMF~\cite{Sawada2013} (Section~\ref{sec:mnmf}),
 MNMF-DP~\cite{Sekiguchi2018} (Section~\ref{sec:mnmf_dp})
 based on the \textit{unconstrained} SCMs,
 where all methods used the MM algorithms 
 (with the IP step for estimating $\Q$)
 described in this paper.
The original FCA~\cite{Duong2010} and FastFCA~\cite{Ito2018}
 denoted by FCA\emalgo\ and FastFCA\emalgo\
 based on the EM algorithms
 (with the FPI step for estimating $\Q$)
 were also tested.
For comparison, ILRMA\cite{Kitamura2016}
 based on the \textit{rank-1} SCMs was tested.
 
The number of sources $N$ was set as $2 \le N \le M$
 except for ILRMA used only in a determined condition $N=M=5$.
The number of iterations was 100.
For the NMF-based source model, 
 the number of bases $K$ was set to $4$, $16$, or $64$.
For the DNN-based source model,
 the latent variables $\Z_1$ with $D=16$ 
 were updated 30 times per iteration
 and the proposal variance $\epsilon$ was set to $10^{-4}$.
The parameters $\bm\theta$ were trained in advance 
 from clean speech data of about 15 hours 
 included in WSJ-0 corpus~\cite{GarofaloJohnandGraffDavidandPaulDougandPallett2007}
 as described in \cite{Leglaive2018}.
More specifically,
 a DNN-based decoder $\bm\sigma^2_{\bm\theta}$ that generates $\X$ from $\Z$
 and a DNN-based encoder that infers $\Z$ from $\X$ 
 were trained jointly 
 in a variational autoencoding manner \cite{Kingma2014}.
The SCM of speech $\G_{1f}$ was initialized as the average of the observed SCMs 
 and the SCMs of noise $\G_{(n \ge 2) f}$ were initialized 
 as the identity matrices.
$\tilde{\G}$ and $\Q$ were initialized 
 with spectral decomposition of $\G$.
$\Z$ was initialized by feeding $\X$ to the encoder.

{\tabcolsep=5.4pt
\renewcommand{\arraystretch}{1.0}
  \begin{table*}[t]
    \caption{The elapsed times per iteration for processing noisy speech signals of 8 [sec].}
    \vspace{-11pt}
    \label{table:processingTime}
    \begin{center}
      \begin{minipage}{\linewidth}
        \subcaption{Elapsed times [sec] on CPU (Intel Xeon W-2145 3.70 GHz)}
        \vspace{-6pt}
        \label{table:processingTime_CPU}
    \begin{tabular}{p{3.5em}|p{1em}|p{8.5em}|p{6.3em}|p{1.3em}p{1.3em}p{1.3em}|p{3.87em}p{3.87em}p{3.87em}|p{3.95em}p{3.95em}p{3.95em}}
            \Hline
            \multicolumn{2}{c|}{Method} 
            & FCA\emalgo\ / FastFCA\emalgo & FCA / FastFCA & \multicolumn{3}{c|}{ILRMA} & \multicolumn{3}{c|}{MNMF / FastMNMF} & \multicolumn{3}{c}{MNMF-DP / FastMNMF-DP} \\
            \hline
            \multicolumn{2}{c|}{\# of bases $K$} & \hfil $-$ \hfil &  \hfil$-$ \hfil &  \hfil4 \hfil & \hfil 16 \hfil &  \hfil64 \hfil &  \hfil4  \hfil& \hfil 16 \hfil &  \hfil64 \hfil &  \hfil4 \hfil & \hfil 16 \hfil &  \hfil64 \hfil \\
            \hline
            \multirow{4}{*}{
            \shortstack{\# of \\ sources \\ $N$}}
            & \hfil 2 \hfil & \hfil  2.1 / 0.49 \hfil  & \hfil  3.3 / 0.43 \hfil  &    \hfil$-$\hfil &\hfil $-$ \hfil& \hfil$-$\hfil   &  4.9 / 0.70   &   5.0 / 0.79   &   5.4 / 1.3   &  \hfil 11 /  1.7  \hfil & \hfil  11 / 1.7  \hfil & \hfil  11 / 1.9\hfil \\
            & \hfil 3 \hfil &  \hfil 2.6 / 0.59 \hfil  & \hfil  4.0 / 0.47 \hfil  &    \hfil$-$\hfil &\hfil $-$ \hfil& \hfil$-$\hfil   &  5.9 / 0.78   &   6.0 / 0.91   &   6.5 / 1.7   &  \hfil 13 /  1.8 \hfil  &  \hfil 13 / 1.8  \hfil &  \hfil 13 / 2.3 \hfil\\
            & \hfil 4 \hfil &  \hfil 3.2 / 0.70 \hfil  &  \hfil 4.7 / 0.56 \hfil  &    \hfil$-$\hfil &\hfil $-$ \hfil& \hfil$-$\hfil   &  6.8 / 0.85   &   7.0 / 1.1    &   7.7 / 2.2   &  \hfil 15 /  1.9 \hfil  &  \hfil 15 / 2.0  \hfil &  \hfil 15 / 2.8 \hfil\\
            & \hfil 5 \hfil &  \hfil 3.7 / 0.81 \hfil  &  \hfil 5.3 / 0.63 \hfil  &  0.53 & 0.62 &\hfil 1.0 \hfil &  7.8 / 1.0   &   8.0 / 1.2     &   8.9 / 2.8   &  \hfil 17 /  2.0 \hfil  &  \hfil 17 / 2.2 \hfil  &  \hfil 17 / 3.4\hfil \\
            \Hline
        \end{tabular}
      \end{minipage}
    \end{center}
    \begin{center}
    \begin{minipage}{\linewidth}
        \vspace{-3pt}
        \subcaption{Elapsed times [decisec] on GPU (NVIDIA GeForce GTX 1080 Ti)}
        \vspace{-6pt}
        \label{table:processingTime_GPU}
        {\tabcolsep=4.85pt
    \begin{tabular}{p{3.8em}|p{1.em}|p{8.58em}|p{6.5em}|p{1.45em}p{1.45em}p{1.45em}|p{4.03em}p{4.03em}p{4.03em}|p{4.1em}p{4.1em}p{4.1em}}
            \Hline
            \multicolumn{2}{c|}{Method} & 
            FCA\emalgo\ / FastFCA\emalgo & FCA / FastFCA & \multicolumn{3}{c|}{ILRMA} & \multicolumn{3}{c|}{MNMF / FastMNMF} & \multicolumn{3}{c}{MNMF-DP / FastMNMF-DP} \\
            \hline
            \multicolumn{2}{c|}{\# of bases $K$} & \hfil $-$ \hfil &  \hfil$-$ \hfil &  \hfil4 \hfil & \hfil 16 \hfil &  \hfil64 \hfil &  \hfil4  \hfil& \hfil 16 \hfil &  \hfil64 \hfil &  \hfil4 \hfil & \hfil 16 \hfil &  \hfil64 \hfil \\
            \hline
            \multirow{4}{*}{
            \shortstack{\# of \\ sources \\ $N$}}
            &\hfil 2 \hfil   & \hfil  1.6 / 0.16 \hfil  & \hfil  2.0 / 0.15 \hfil  &    \hfil$-$\hfil &\hfil $-$ \hfil& \hfil$-$\hfil  &   3.0 / 0.38   &   3.0 / 0.55   &   3.2 / 1.2   & \hfil  7.0 / 0.90 \hfil& \hfil7.0 / 0.98 \hfil  & \hfil 7.1 / 1.3 \hfil\\
            &\hfil 3 \hfil  &  \hfil 2.3 / 0.19 \hfil  & \hfil  2.8 / 0.17 \hfil  &    \hfil$-$\hfil &\hfil $-$ \hfil& \hfil$-$\hfil  &   4.2 / 0.43   &   4.2 / 0.68   &   4.5 / 1.7   &   \hfil9.3 / 0.94\hfil & \hfil9.3 / 1.1 \hfil  & \hfil  9.5 / 1.8\hfil \\
            & \hfil 4  \hfil & \hfil  3.0 / 0.22 \hfil  &  \hfil 3.6 / 0.17 \hfil &    \hfil$-$\hfil &\hfil $-$ \hfil& \hfil$-$\hfil  &   5.3 / 0.46   &   5.4 / 0.81   &   5.7 / 2.2   &  \hfil 12 / 0.99 \hfil & \hfil12 / 1.2  \hfil &  \hfil 12 / 2.3\hfil \\
            & \hfil 5 \hfil  &  \hfil 3.7 / 0.25 \hfil  &  \hfil 4.5 / 0.19 \hfil  & 0.52 & 0.61 &\hfil 1.0\hfil &   6.6 / 0.51   &   6.7 / 0.94   &   7.1 / 2.7   & \hfil  14 / 1.0 \hfil  & \hfil14 / 1.4 \hfil  &   \hfil14 / 2.8\hfil \\
            \Hline
        \end{tabular}
      }
    \end{minipage}
  \end{center}
  \end{table*}
\renewcommand{\arraystretch}{1.0}
}

{\tabcolsep=3.85pt
\renewcommand{\arraystretch}{1.0}
  \begin{table*}[!t]
  \vspace{-11pt}
  \caption{The average SDRs [dB] for 100 noisy speech signals.}
  \vspace{-7pt}
  \label{table:SDR_NK}
  \begin{center}
      \begin{minipage}{\linewidth}
      \vspace{-5pt}
    \begin{tabular}{p{4em}|p{1.4em}|p{8.9em}|p{6.7em}|p{1.7em}p{1.7em}p{1.7em}|p{4.3em}p{4.3em}p{4.3em}|p{4.3em}p{4.3em}p{4.3em}}
    \Hline
        \multicolumn{2}{c|}{Method} & 
        FCA\emalgo\ / FastFCA\emalgo & FCA / FastFCA & \multicolumn{3}{c|}{ILRMA} & \multicolumn{3}{c|}{MNMF / FastMNMF} & \multicolumn{3}{c}{MNMF-DP / FastMNMF-DP} \\
        \hline
        \multicolumn{2}{c|}{\# of bases $K$} & \hfil $-$ \hfil &  \hfil$-$ \hfil &  \hfil4 \hfil & \hfil 16 \hfil &  \hfil64 \hfil &  \hfil4  \hfil& \hfil 16 \hfil &  \hfil64 \hfil &  \hfil4 \hfil & \hfil 16 \hfil &  \hfil64 \hfil \\
        \hline
        \multirow{4}{*}{
        \shortstack{\# of \\ sources \\ $N$}}
        &\hfil 2\hfil &  \hfil 8.9 / 10.3 \hfil  &   \hfil 8.6 / 10.5 \hfil  &  \hfil$-$\hfil &\hfil $-$ \hfil& \hfil$-$\hfil &  11.4 / 15.3    &    11.1 / 15.6    &   10.5 / 15.1   &   17.5 / 17.5   &   18.1 / 18.2   &   18.5 / 18.6   \\
        & \hfil3 \hfil&  \hfil 9.1 / 10.8 \hfil & \hfil  8.8 / 11.1 \hfil  &   \hfil$-$\hfil &\hfil $-$ \hfil& \hfil$-$\hfil  & 12.3 / 16.1    &    12.0 / 16.4    &   11.3 / 15.8   &   18.0 / 18.3   &   18.4 / 18.6   &   18.6 / 18.8   \\
        & \hfil4 \hfil& \hfil  9.3 / 11.0 \hfil  &  \hfil 8.8 / 11.6 \hfil  &  \hfil$-$\hfil &\hfil $-$ \hfil& \hfil$-$\hfil  & 13.0 / 16.2    &    12.7 / 16.7    &   11.9 / 16.1   &   18.0 / 18.4   &   18.4 / \textbf{18.9}   &   18.4 / \textbf{18.9}   \\
        & \hfil5\hfil & \hfil  9.4 / 11.1 \hfil  & \hfil  8.9 / 11.9 \hfil  &  15.1 & 15.1 & 14.9 & 13.2 / 16.4    &    13.1 / 16.8    &   12.4 / 16.3   &   18.2 / 18.6   &   18.2 / 18.8   &   18.1 / 18.8   \\
        \Hline
      \end{tabular}
      \end{minipage}
  \end{center}
  \vspace{-12.3pt}
  \end{table*}
\renewcommand{\arraystretch}{1.0}
}

\subsection{Experimental Results}

Tables \ref{table:processingTime}-\subref{table:processingTime_CPU} 
 and \ref{table:processingTime}-\subref{table:processingTime_GPU} 
 list the elapsed times per iteration 
 and Table \ref{table:SDR_NK} lists the average SDRs.
FastFCA slightly outperformed FastFCA\emalgo~\cite{Ito2018} in all measures
 because the IP method and the FPI method
 calculate the matrix inversion only once and twice, respectively, for updating $\Q$,
 and the MM algorithm converges faster than the EM algorithm.
FastFCA, FastMNMF, and FastMNMF-DP were an order of magnitude faster
 and performed as well as or even better than their original versions.
In general, more than any two positive definite matrices
 cannot be exactly jointly diagonalized.
If $N \ge 3$, 
 the fast versions are thus inferior to the original versions
 in terms of the DOF,
 but the restriction of the DOF of the spatial model
 was proved to be effective for avoiding bad local optima.
If $N = 2$, the DOFs of the fast versions
 are exactly the same as those of the original versions in theory
 as described in stereo FastFCA \cite{Ito2018a},
 but the fast versions were less sensitive 
 to the initialization in our experiment.
One reason would be that 
 while only the SCM of speech $\G_{1f}$ was initialized to a reasonable value in the original versions,
 the initialization of $\Q_f$ based on $\G_{1f}$
 contributed to initializing the SCM of noise in the fast versions.
When $N=5$ and $K=4$ (the best condition for ILRMA),
 FastMNMF was as fast as and outperformed ILRMA.

\section{Conclusion}

This paper presented a full-rank spatial model
 based on the jointly diagonalizable SCMs of sound sources
 and its application to existing methods such as FCA, MNMF, and MNMF-DP.
For such fast versions,
 we proposed a general convergence-guaranteed MM algorithm
 that uses the IP method for estimating the SCMs.
We experimentally showed that 
 our approach is effective for improving
 both the separation performance and computational efficiency.
One important direction 
 is to develop online FastMNMF-DP for real-time noisy speech recognition
 because the real-time factor of FastMNMF-DP could be less than 1.
We also plan to simultaneously consider 
 the jointly diagonalizable \textit{full-rank} 
 spatial and frequency covariance matrices of sound sources
 as suggested in \cite{Yoshii2018}.\\

 \bibliography{eusipco-2019-sekiguchi}
\bibliographystyle{IEEEtran}

\end{document}